%% file: cirq_opt_quantum.tex
\documentclass[a4paper,twocolumn,11pt,accepted=2022-09-01]{quantumarticle}
\pdfoutput=1
\usepackage[utf8]{inputenc}
\usepackage[english]{babel}
\usepackage[T1]{fontenc}
\usepackage{amsmath}
\usepackage{hyperref}

\usepackage{tikz}
\usepackage{lipsum}

\usepackage{amssymb}
\usepackage{comment}
\usepackage{algorithm}
\usepackage{algorithmic}
\usepackage{xcolor}
\usepackage{braket}
\usepackage[numbers]{natbib}

\usepackage{qcircuit}
\usepackage{adjustbox}

\RequirePackage{xspace}
\newcommand{\tket}{\ensuremath{\text{t}\hspace{-0.5mm}\ket{\text{ket}}}\xspace}
\newcommand{\myopt}{{\sc Aqcel}\xspace}
\newcommand{\slow}{\ensuremath{s_\epsilon^\text{dyn}}\xspace}

\newcommand{\sfrac}{\ensuremath{s_\epsilon^f}\xspace}
\newcommand{\porig}{\ensuremath{p^{\text{orig}}}}
\newcommand{\popt}{\ensuremath{p^{\text{opt}}}}
\newcommand{\Fsim}{\ensuremath{F_{\text{sim}}}}
\newcommand{\Fmeas}{\ensuremath{F_{\text{meas}}}}
\newcommand{\Fqst}{\ensuremath{F_{\text{QST}}}}

\begin{document}

\title{Initial-State Dependent Optimization of Controlled Gate Operations with Quantum Computer}

\author{Wonho Jang}
\email{jang@icepp.s.u-tokyo.ac.jp}
\affiliation{Department of Physics, The University of Tokyo, 7-3-1 Hongo, Bunkyo-ku, Tokyo 113-0033, Japan}
\author{Koji Terashi}
\email{koji.terashi@cern.ch}
\orcid{0000-0001-6520-8070}
\affiliation{International Center for Elementary Particle Physics (ICEPP), The University of Tokyo, 7-3-1 Hongo, Bunkyo-ku, Tokyo 113-0033, Japan}
\author{Masahiko Saito}
\affiliation{International Center for Elementary Particle Physics (ICEPP), The University of Tokyo, 7-3-1 Hongo, Bunkyo-ku, Tokyo 113-0033, Japan}
\author{Christian W. Bauer}
\affiliation{Physics Division, Lawrence Berkeley National Laboratory, Berkeley, CA 94720, USA}
\author{Benjamin Nachman}
\affiliation{Physics Division, Lawrence Berkeley National Laboratory, Berkeley, CA 94720, USA}
\author{Yutaro Iiyama}
\affiliation{International Center for Elementary Particle Physics (ICEPP), The University of Tokyo, 7-3-1 Hongo, Bunkyo-ku, Tokyo 113-0033, Japan}
\author{Ryunosuke Okubo}
\affiliation{Department of Physics, The University of Tokyo, 7-3-1 Hongo, Bunkyo-ku, Tokyo 113-0033, Japan}
\author{Ryu Sawada}
\affiliation{International Center for Elementary Particle Physics (ICEPP), The University of Tokyo, 7-3-1 Hongo, Bunkyo-ku, Tokyo 113-0033, Japan}

\maketitle

\begin{abstract}
There is no unique way to encode a quantum algorithm into a quantum circuit.  With limited qubit counts, connectivity, and coherence times, a quantum circuit optimization is essential to make the best use of near-term quantum devices.  We introduce a new circuit optimizer called {\sc Aqcel}, which aims to remove redundant controlled operations from controlled gates,  depending on initial states of the circuit.
Especially, the {\sc Aqcel} can remove unnecessary qubit controls from multi-controlled gates in polynomial computational resources, even when all the relevant qubits are entangled, by identifying zero-amplitude computational basis states using a quantum computer.
As a benchmark, the {\sc Aqcel} is deployed on a quantum algorithm designed to model final state radiation in high energy physics. For this benchmark, we have demonstrated that the {\sc Aqcel}-optimized circuit can produce equivalent final states with much smaller number of gates. Moreover, when deploying {\sc Aqcel} with a noisy intermediate scale quantum computer, it efficiently produces a quantum circuit that approximates the original circuit with high fidelity by truncating low-amplitude computational basis states below certain thresholds. Our technique is useful for a wide variety of quantum algorithms, opening up new possibilities to further simplify quantum circuits to be more effective for real devices.
\end{abstract}

\section{Introduction}
\label{sec:intro}
\input{intro}

\section{\myopt optimization protocol}
\label{sec:algo}
\input{algorithm}

\section{Application to quantum algorithm}
\label{sec:application}
\input{application}

\section{Discussion}
\label{sec:discussion}
\input{discussion}

\section{Conclusion and outlook}
\label{sec:conclusion}
\input{conclusion}

\begin{acknowledgements}
\input{acknowledgements}
\end{acknowledgements}

\appendix
\input{appendix}

\bibliographystyle{quantum}
\bibliography{cirq_opt_quantum}

\end{document}

%% file: intro.tex
Recent technology advances have resulted in a variety of universal quantum computers that are being used to implement quantum algorithms.  However, these noisy-intermediate-scale quantum (NISQ) devices~\cite{Preskill_2018} may not have sufficient qubit counts, qubit connectivity and capability to stay coherent for the entirety of operations in a particular algorithm implementation.  Despite these challenges, a variety of applications have emerged across science and industry.  For example, there are many promising studies in experimental and theoretical high energy physics (HEP) for exploiting quantum computers.  These studies include event classification~\cite{Mott:2017xdb,Zlokapa_2020,Chan:2019zwk,terashi2020event,Guan:2020bdl,belis2021higgs}, reconstructions of charged particle trajectories~\cite{Zlokapa:2019tkn,Tuysuz:2020ocw,Shapoval:2019txi,Bapst:2019llh} and physics objects~\cite{Wei:2019rqy,Das:2019hrw}, unfolding measured distributions~\cite{Cormier:2019kcq} as well as simulation of multi-particle emission processes~\cite{Bauer:2019qx,Nachman_2021}.  A common feature of all of these algorithms is that only simplified versions can be run on existing hardware due to the limitations mentioned above.

There are generically two strategies for improving the performance of NISQ computers to execute existing quantum algorithms. One strategy is to mitigate errors through active or passive modifications to the quantum state preparation and measurement protocols. For example, readout errors can be mitigated through post-processing steps~\cite{Bauer:2019uf,chen_detector_2019,dewes_characterization_2012,geller_efficient_2020,geller_rigorous_2020,2010.07496} and gate errors can be mitigated by systematically enlarging errors before extrapolating to zero error~\cite{Dumitrescu:2018,PhysRevX.8.031027,PhysRevLett.119.180509,Kandala:2019,PhysRevA.102.012426,Otten_2019}.  A complementary strategy to error mitigation is quantum compilation. There is no unique way to encode a quantum algorithm into a set of gates, and certain realizations of an algorithm may be better suited for a given quantum device. Widely used tools are Qiskit~\cite{Qiskit} and t|ket$\rangle$~\cite{Sivarajah_2020}, which contain a variety of architecture-agnostic and architecture-specific routines. There are also a variety of other toolkits for circuit optimization, including hardware-specific packages for quantum circuits~\cite{H_ner_2018,Green_2013,JavadiAbhari_2015,Svore_2018,Killoran_2019,Qiskit,smith2016practical,Steiger_2018,quantum_ai_team_and_collaborators_2020_4062499,mccaskey2018language,murali2019fullstack,robert_s_smith_2020_3677537,Nam_2018,venturelli2019quantum,murali2019noiseadaptive,Murali_2020,Peterson_2020,Leung_2017,Gokhale_2019,liu2020relaxed}.

Among the gates used for an algorithm encoding, multi-controlled gates are significant error sources because they result in many CNOT gates after the decomposition~\cite{PhysRevA.52.3457}, and also require SWAP gates to fit within limited qubit topology. The costs to implement multi-controlled gates can be reduced by using relative phase Toffli gates~\cite{PhysRevA.93.022311}, ancilla qubits~\cite{addline} or qutrits~\cite{Asymptotic, 5954250, galda2021implementing, inada2021measurementfree, Wang, Ralph, Kiktenko} in the implementation. 

An alternative approach for reducing the costs is to remove unnecessary qubit controls. The reversible circuit synthesis can reduce redundant qubit controls while maintaining the equivalence of a quantum circuit before and after the optimization~\cite{Zhong2006UsingCF}. 
Previous work with reversible circuit synthesis has largely focused on circuits composed of CNOT gates~\cite{Ketan_RevCircuit}. The circuit synthesis is extended later to more general quantum circuits, e.g., those composed of CNOTs and $Z$-basis rotation gates~\cite{Amy_2018}.
It is possible to remove, beyond reversible circuit synthesis, more unnecessary controlled operations if we consider maintaining the equivalence of the final state. Imagine that there is a $n$-qubit quantum circuit designed to work with different initial states and it is executed with a given initial state such as $\ket{0}^{\otimes n}$. In this case, the circuit will reach only a selected set of intermediate states and some operations may become trivial. Such initial-state dependent circuit optimization may find more rooms for optimization if the equivalence of the final state, not the circuit itself, is preserved. Thus, it will enable more aggressive reduction of unnecessary controlled operations than initial-state independent circuit optimization.

There are two main approaches for initial-state dependent circuit optimization. The first one is a continuous optimization that trains an ansatz with parameters~\cite{Khatri2019quantumassisted, energy_minimisation}. The second one is a discretized optimization in which some controlled operations are removed from a gate if the quantum state satisfies a specific condition at the point where the gate is operated~\cite{liu2020relaxed,Coecke_2011,Duncan_2020}. We focus on the latter in this paper.
Among existing discretized optimization protocols that account for initial states, the Relaxed Peephole Optimization (RPO)~\cite{liu2020relaxed} reduces controlled operations when the qubits in the X- or Z-basis states are used as control qubits or a target qubit of a controlled gate. This protocol, however, cannot remove qubit controls in the case where all relevant qubits are entangled.
The ZX-calculus~\cite{Coecke_2011,Duncan_2020} exploits a copy-rule for removing a qubit control from CNOT gate when the control qubit state is $\ket{1}$. This leads to an initial-state dependent circuit optimization, but it cannot remove all qubit controls within polynomial complexity~\footnote{In fact, the ZX-calculus is complete in the formal logic sense of the word, such that one can always prove that all unnecessary qubit controls can be removed using rules of the ZX-calculus~\cite{ZXcomp}. However, in general this scheme requires exponential resources. Nevertheless, the ZX-calculus is still incredibly powerful and underlies many of the optimization techniques of quantum transpilers.}.

The novel optimization protocol proposed in this paper has three distinct features. First, there is the ability of removing redundant qubit controls no matter whether all the relevant qubits are entangled or not. Second, the identification of zero- or low-amplitude computational basis states using a quantum computer allows one to obtain bitstrings in polynomial time, otherwise exponential resources are required in the classical calculations. Third, the decomposition of multi-controlled $U$ gates into Toffoli gates and singly-controlled $U$ gates enables us to perform the search of all unnecessary qubit controls in polynomial resources. This new optimization protocol also serves a new efficient method to approximate quantum circuits in the NISQ era by truncating low-amplitude computational basis states that do not contribute significantly to the final state.

This optimization protocol is called \textsc{Aqcel} (and pronounced ``excel'') for \textit{Advancing Quantum Circuit by \textsc{icEpp} and \textsc{Lbnl}}. To demonstrate the effectiveness of the \myopt protocol, we will use a quantum algorithm that models a \textit{parton shower}~\cite{Nachman_2021}. This algorithm provides a useful benchmark because it is designed to work with different initial states corresponding to different initial particles, meaning that the quantum circuits have redundancy for a specific initial state.

This paper is organized as follows.  Section~\ref{sec:algo} provides an overview of the \myopt protocol.  The application of this protocol to the HEP example is presented in Sec.~\ref{sec:application}.  Following a brief discussion about the applicability and future extensions of the protocol in Sec.~\ref{sec:discussion}, the paper concludes in Sec.~\ref{sec:conclusion}.

%% file: algorithm.tex
First, we summarize the concept of removing redundant controlled operations from controlled gates depending on initial states of a quantum circuit. Then, the \myopt protocol of removing redundant qubit controls and the methods for executing the whole optimization in polynomial resources are described.

\subsection{Basic idea of redundant controlled operations removal}
\label{subsec:basic_idea}
A controlled gate performs a different operation depending on the quantum state at the point where the gate is applied. Let $m$ be the number of control qubits of this gate. Consider expanding the state of the full system $\ket{\psi}$ into a superposition of computational basis states as
\begin{equation}\label{eq:state_decomposition}
  \ket{\psi} = \sum_{j,k} c_{j,k} \ket{j}_{\text{ctl}} \otimes \ket{k},
\end{equation}
where $\ket{\cdot}_{\text{ctl}}$ denotes the state of the control qubits, while the unlabeled ket corresponds to the rest of the system. 
We write the states as integers with $0 \leq j \leq 2^m-1$ and $0 \leq k \leq 2^{n-m} - 1$. We assume that the controlled gate is applied to computational basis states whose bitstrings on all control qubits are  $1$, which corresponds to the state $\ket{j}_{\text{ctl}} = \ket{11 \cdots 1} = \ket{2^{m}-1}_{\text{ctl}}$. This allows one to classify the state of the system into three general classes using the amplitudes $c_{j,k}$:
\begin{description}
\item[\textbf{Triggering} :] $c_{j,k} \neq 0$ if and only if $j = 2^{m}-1$. The controlled operation of the gate in question is applied for all computational bases in the superposition. \vspace{-2mm}
\item[\textbf{Non-triggering} :] $c_{2^{m}-1,k} = 0$ for all $k$. The controlled operation is never applied.\vspace{-2mm}
\item[\textbf{Undetermined} :] The state is neither triggering nor non-triggering.
\end{description}

A circuit containing triggering or non-triggering controlled gates can be simplified by removing all controls (triggering case) or by eliminating the gates entirely (non-triggering case). While an
undetermined single-qubit controlled gate cannot be simplified under the current scheme, an undetermined multi-qubit controlled gate can be by removing the controls on some of the qubits, if the state of the system satisfies the condition, and that is our interest.

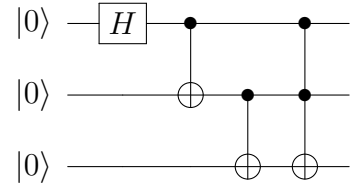
\begin{figure}[h!]
\centering
\leavevmode
\large
\Qcircuit @C=1em @R=1em @!R{
&&&\lstick{\ket{0}}  &\gate{H} & \ctrl{1} & \qw      & \ctrl{1} &\qw  \\
&&&\lstick{\ket{0}}  & \qw     &\targ     & \ctrl{1} & \ctrl{1} &\qw  \\
&&&\lstick{\ket{0}}  & \qw     & \qw      &\targ     & \targ    &\qw  \\
}
\caption{A quantum circuit in $\ket{0}^{\otimes 3}$ initial state. The control and target qubits for the Toffoli gate are entangled because the Hadamard and CNOT gates create the GHZ state.}
\label{fig:sample}
\end{figure}
As an example of this concept, consider the simple circuit in Fig.~\ref{fig:sample} composed of three qubits in $\ket{0}^{\otimes 3}$ initial state. 
At the Toffoli gate, the quantum state is in the superposition of $\ket{000}$ and $\ket{111}$, which is a Greenberger-Horne-Zeilinger state where the qubits are maximally entangled. This is an undetermined state for the Toffoli gate. Moreover, all the control qubits and the target qubit are entangled, which is a difficult case for the qubit control reduction. However, the \myopt can remove one of the two qubit controls from the Toffoli, replacing it with a CNOT gate controlled only by the remaining one.

\subsection{General conditions to eliminate qubit controls}
\label{subsec:qubit_control}

Given a multi-qubit controlled-$U$gate with $m$ control qubits, denoted by $C^{m}[U]$, and a system in an undetermined state $\ket{\psi}$ defined in Sec.~\ref{subsec:basic_idea}, we can derive general conditions for a part of the controlled operations to be removed, as follows.

Let $x$ ($<m$) be the number of controls to be removed. Without loss of generality, the decomposition of $\ket{\psi}$ can be rewritten as
\begin{equation}
  \ket{\psi} = \sum_{i, l, k} \tilde{c}_{i, l, k} \ket{i}_{\text{ctl}'} \otimes \ket{l}_{\text{free}} \otimes \ket{k},
\end{equation}
where $\ket{\cdot}_{\text{ctl}'}$ and $\ket{\cdot}_{\text{free}}$ are the states of the $m-x$
remaining control qubits and the $x$ qubits from which the controls are removed. From Eq.~\eqref{eq:state_decomposition},
\begin{equation}\label{eq:basis_breakdown}
  \ket{i}_{\text{ctl}'} \otimes \ket{l}_{\text{free}} = \ket{2^{x} i + l}_{\text{ctl}},
\end{equation}
and therefore
\begin{equation}\label{eq:coeff_identity}
  \tilde{c}_{i, l, k} = c_{2^{x}i + l, k}.
\end{equation}

Applying the original controlled gate to $\ket{\psi}$ yields
\begin{multline}\label{eq:controlled_original}
  C^{m}[U] \ket{\psi} = \sum_{j=0}^{2^{m} - 2} \sum_{k} c_{j, k} \ket{j} \ket{k} + \\
  \sum_{k} c_{2^{m} - 1, k} \ket{2^{m} - 1} U \ket{k},
\end{multline}
where ket subscripts and the tensor product symbols are omitted for simplicity.
In contrast, a new gate with fewer controls gives
\begin{multline}\label{eq:controlled_reduced}
  C^{m-x}[U] \ket{\psi} = \sum_{i=0}^{2^{m-x} - 2} \sum_{l,k} \tilde{c}_{i, l, k} \ket{i} \ket{l} \ket{k} + \\
  \sum_{l, k} \tilde{c}_{2^{m-x} - 1, l, k} \ket{2^{m-x} - 1} \ket{l} U \ket{k}.
\end{multline}

For the removal of $x$ qubit controls to be allowed, the right hand sides of
Eqs.~\eqref{eq:controlled_original} and \eqref{eq:controlled_reduced} must be identical. This requires
\begin{multline}\label{eq:state_identity}
  \sum_{l=0}^{2^{x} - 2} \sum_{k} \tilde{c}_{2^{m-x} - 1, l, k} \ket{2^{m - x} - 1} \ket{l} U \ket{k} = \\
  \sum_{l=0}^{2^{x} - 2} \sum_{k} c_{2^{m} - 2^{x} + l, k} \ket{2^{m} - 2^{x} + l} \ket{k}.
\end{multline}
Denoting
\begin{equation}
  U \ket{k} = \sum_{k'} u_{kk'} \ket{k'}
\end{equation}
and recalling Eq.~\eqref{eq:basis_breakdown}, Eq.~\eqref{eq:coeff_identity}, Eq.~\eqref{eq:state_identity} implies (replacing $k' \leftrightarrow k$ on the left hand side)
\begin{multline}
  \sum_{l=0}^{2^{x} - 2} \sum_{k,k'} \tilde{c}_{2^{m-x} - 1, l, k'} u_{k'k} \ket{2^{m - x} - 1} \ket{l} \ket{k} = \\
  \sum_{l=0}^{2^{x} - 2} \sum_{k} \tilde{c}_{2^{m-x} - 1, l, k} \ket{2^{m - x} - 1} \ket{l} \ket{k}.
\end{multline}
Then, we have
\begin{multline}\label{eq:the_gate_condition}
  \sum_{k'} \tilde{c}_{2^{m - x} - 1, l, k'} u_{k'k} = \tilde{c}_{2^{m - x} - 1, l, k} \\
  \quad \forall l \in \{0, 1, ..., 2^x - 2\}, k.
\end{multline}
Eq.~\eqref{eq:the_gate_condition} holds if the row vector $\{\tilde{c}_{2^{m - x} - 1, l, k}\}_{k}$
is an eigenvector of the matrix $u$ with eigenvalue 1 under right multiplication for $0 \leq l \leq 2^x - 2$, or if
$\tilde{c}_{2^{m - x} - 1, l, k} = 0$ for $0 \leq l \leq 2^x - 2$ and all $k$.

Since the cost of exactly computing the complex amplitudes of the quantum state is exponential, in \myopt
we only consider this second condition:
\begin{equation}\label{eq:aqcel_condition}
  \tilde{c}_{2^{m - x} - 1, l, k} = 0 \quad \forall l \in \{0, 1, ..., 2^x - 2\}, k.
\end{equation}
This removal of redundant qubit controls therefore requires us to find out if $\ket{\psi}$ satisfies Eq.~\eqref{eq:aqcel_condition} for each controlled gate.

In the previous optimization based on RPO, if the target qubit is assumed to be entangled, the condition for the removal of qubit controls from $C^{m}[U]$ is that control qubits which will be removed must be $\ket{1}$ in a pure state. The condition can be written as
\begin{equation}\label{eq:rpo_condition}
  \tilde{c}_{i, l, k} = 0 \quad \forall l \in \{0, 1, ..., 2^x - 2\}, i, k.
\end{equation}
Eq.~\eqref{eq:rpo_condition} is a sufficient condition of Eq.~\eqref{eq:aqcel_condition}, which means that the \myopt has  more rooms for removal of unnecessary qubit controls. A difference between Eq.~\eqref{eq:aqcel_condition} and Eq.~\eqref{eq:rpo_condition} resides in whether one can remove unnecessary qubit controls from entangled control qubits or not, under the condition that target qubit is also entangled. 

\subsection{Identification of computational basis states}
\label{subsec:comp-base}
In general, a circuit consisting of $n$ qubits creates a quantum state described by a superposition of all of the $2^n$ computational basis states. However, it is rather common that a specific circuit produces a quantum state where only a subset of the computational basis states has nonzero amplitudes. Moreover, the number of nonzero-amplitude basis states depends on the initial state. This is why the three classes of the states on control qubits arise.

As mentioned in Sec.~\ref{subsec:qubit_control}, we have to find out if $\ket{\psi}$ satisfies Eq.~\eqref{eq:aqcel_condition}. It can be written down as
\begin{equation}\label{eq:bit_possibilities}
    \sum_{k} |\tilde{c}_{2^{m - x} - 1, l, k}|^2 =0 \quad \forall l \in \{0, 1, ..., 2^x - 2\}.
\end{equation}
Eq.~\eqref{eq:bit_possibilities} requires that there is no computational basis state whose bitstring on all control qubits of $C^{m-x}[U]$ is $\ket{11 \cdots 1}$, except when the bitstring on the removed $x$ control qubits is also $\ket{11 \cdots 1}$. In other words, there should be no bitstring by which $C^{m}[U]$ is not triggered but $C^{m-x}[U]$ is. This can be verified by the identification of bitstrings on all control qubits of $C^{m}[U]$.

The possible bitstrings on control qubits at each controlled gate can be determined either through a classical simulation or by measuring the control qubits by a quantum computer repeatedly. In the case of a classical simulation, one performs the full calculation of the amplitudes.
When instead the quantum measurements are used, the circuit is truncated right before the controlled gate in question, and the control qubits are measured repeatedly at the truncation point. Finiteness of the relevant amplitudes can be inferred from the distribution of the obtained bitstrings, albeit within the statistical uncertainty of the measurements~\footnote{When the number of measurements is not enough for a part of low-amplitude computational basis states, the \myopt approximates the original circuit to a simpler circuit, but this usually loses the identity of the final state. There is a similar effect when low-amplitude computational basis states below thresholds are truncated, as described later.}.

A few notes should be taken on the computational costs of the two methods. Consider an $n$-qubit circuit with $N$ controlled gates. 
A classical simulation of the state vector before a given controlled gate has an exponential scaling in the number of qubits and requires
$\mathcal{O}(2^n)$ computations.
On the other hand, measuring $m$ control qubits $M$ times on each controlled gates by a quantum computer only requires ${\mathcal O}(MN^2+mMN)$ operations which scales only polynomially with the number of qubits.
More details on the estimates of the computational resource necessary for the identification of computational basis states are described in Appendix~\ref{app:comp_resource}.

Note that for noisy quantum computers the measurements of the bitstrings will not be exact due to hardware noise. The list of observed bitstrings would contain contributions from errors on the preceding gates and the measurement itself. In \myopt, we obtain the calibration matrix for the control qubits (with 8192 shots per measurement) using Qiskit Ignis API~\cite{Qiskit}. The matrix is then applied to the observed distribution with a least-squares fitting approach.
To deal with remaining error contributions after the measurement error mitigation, we opt to ignore the observed bitstrings with occurrence below certain thresholds~\footnote{In the actual implementation, the threshold of 0.005 is always applied to suppress contributions from imperfect measurement error mitigation.}. Once such a threshold has been decided, the number of measurements required has to be large enough for the statistical uncertainty to be smaller than this threshold~\footnote{This is justified under the assumption that the residual gate errors act as a perturbation, inserting spurious computational basis states with small amplitudes into the superposition of the system.}.

In order to choose the thresholds, we consider gate errors in the single-qubit gates and CNOT gates~\footnote{The reported error rates at the time of the experiment, measured during the preceding calibration run of the hardware, are used for the threshold calculation.}.
Let the single-qubit gate and CNOT error rates be $\epsilon_{\mathrm{u}}^{(i)}$ and $\epsilon_{\mathrm{cx}}^{(i,j)}$, respectively, with $i$ and $j$ indicating qubits that the gates act on.
We can approximate the probabilities, $p_{\mathrm{u}}$ and $p_{\mathrm{cx}}$, of measuring the bitstrings without any single-qubit gate or CNOT gate errors occurring anywhere in the circuit by performing qubit-wise (index-dependent) multiplications of the error rates:
\begin{align}
p_{\mathrm{u}} &= \prod_{i} \left(1-\epsilon_{\mathrm{u}}^{(i)}\right)^{n_{\mathrm{u}}^{(i)}},\\
p_{\mathrm{cx}} &= \prod_{i\neq j} \left(1-\epsilon_{\mathrm{cx}}^{(i,j)}\right)^{n_{\mathrm{cx}}^{(i,j)}},
\end{align}
where $n_{\mathrm{u}}^{(i)}$ and $n_{\mathrm{cx}}^{(i,j)}$ are the numbers of single-qubit gates and CNOT gates acting on the corresponding qubits, respectively. The probability $p_\epsilon$ of measuring the bitstirngs with at least one gate error occurring anywhere in the circuit is
\begin{align}
p_\epsilon &= 1-p_{\mathrm{u}}p_{\mathrm{cx}}
\nonumber\\
&\sim N_{\mathrm{cx}} \epsilon_{\mathrm{cx}}.
\end{align}
In the last approximation, we have assumed that all CNOT errors are equal, and much larger than single-qubit gate errors but still much smaller than one: $\epsilon_{\mathrm{u}}^{(i)} \ll \epsilon_{\mathrm{cx}}^{(i,j)} = \epsilon_{\mathrm{cx}} \ll 1$.
From the $p_\epsilon$, the first threshold is chosen to be
\begin{equation}
\slow := p_\epsilon / 2^m,
\end{equation}
where $m$ is the number of the measured control qubits. This choice of dynamical threshold is motivated by assuming that the quantum errors would result in a uniform distribution of all possible bitstrings according to a depolarizing error model. It should be noted that the $p_\epsilon$ increases as the circuit execution proceeds because the $p_\epsilon$ accounts for the error rates from all the preceding gates in the circuit. As an alternative strategy to the dynamical threshold, we also examine the static thresholds, \sfrac, that are kept constant throughout the circuit, with the values between 0.005 and 0.3.

Discarding all bitstrings with occurrence under certain thresholds usually modifies the final state of the optimized circuit from one of the original circuit.
On the other hand, applying certain thresholds will leave high-amplitude computational basis states, while rejecting low-amplitude computational basis states which do not contribute to the final result meaningfully~\footnote{Here the low-amplitude computational basis states mean a part of the original computational basis states composing of the ideal quantum state of interest, not those from quantum noise.}. Thus, with the proper thresholds, one can produce a quantum circuit which well approximates the original circuit by removing unimportant qubit controls that trigger on low-amplitude computational basis states.
In other words, the actual threshold of \myopt should be selected by considering the trade-off between the noise resilience and the identity of the final state to the original ideal state.

\subsection{Elimination of redundant controlled operations}
\label{subsec:ctrl-gate}
Once the nonzero-amplitude computational basis states are identified at each controlled gate, the next step is to figure out which qubit controls can be removed using Eq.~\eqref{eq:aqcel_condition}. 
The computational cost of determining the removal of redundant qubit controls would be at most ${\mathcal O}(Mm4^mN)$ (Appendix~\ref{app:comp_resource}), which scales exponentially with $m$, the number of control qubits ($N$ is the number of multi-qubit controlled gates in the circuit). To avoid this, a generic multi-qubit controlled gate should be decomposed into controlled gates with small, fixed number of control qubits. 
An arbitrary multi-qubit controlled-$U$ gate with $m$ control qubits can be decomposed into $\mathcal{O}(m)$ Toffoli and controlled-$U$ gates~\cite{PhysRevA.52.3457}. Besides, these Toffoli gates can be replaced with relative phase implementation of Toffoli gates (referred to as just ``Toffoli'' hereafter)~\cite{PhysRevA.93.022311}, which reduces the CNOT counts from 6 (in the regular Toffoli decomposition) to 3. Therefore, in the \myopt scheme, we assume that all controlled gates in a quantum circuit are reduced to Toffoli gates denoted as $C^2[X]$ and singly-controlled unitary operations denoted as $C[U]$. This results in a significant reduction of computational cost of the decision of all redundant qubit controls, because all controlled gates have either 1 or 2 control qubits. However, when controlled gates are decomposed, \myopt would lose a part of opportunity to remove redundant qubit controls. More details about the decomposition are described in Appendix~\ref{app:decomposition}.
Since a multi-qubit controlled gate is decomposed into a set of Toffoli gates and its mirror for uncomputation except the central gate, the optimization of controlled operations for the set of Toffoli gates can be applied to the uncomputation part as well.

For a $n$-qubit circuit composed of $N$ multi-qubit controlled-$U$ gates, each having at most $n$ control qubits, this decomposition results in at most $nN$ controlled gates. With $nN$ gates, the cost for identifying computational basis states (Sec.~\ref{subsec:comp-base}) increases up to $\mathcal{O}(n^2N^2M)$ when measuring with a quantum computer. However, the cost for removing unnecessary qubit controls improves from the above exponential scaling to $\mathcal{O}(MnN)$. More details about the resource scaling are given in Appendix~\ref{app:comp_resource}.

After the decomposition, a $C[U]$ gate can be a single unitary $U$ gate if the probability of observing $\ket{1}$ of the control qubit is 1, or removed if the probability is 0. In all other cases, the $C[U]$ gate is kept.
For a $C^2[X]$ gate, the similar control reduction can be performed with the probabilities of the four possible states $\ket{00}$, $\ket{01}$, $\ket{10}$ and $\ket{11}$. If the probability of the state $\ket{01}$ ($\ket{10}$) is zero, one can eliminate the first (second) control from the $C^2[X]$ gate (see Eq.~\eqref{eq:aqcel_condition}). The following pseudocode is the full algorithm for redundant controlled operations removal.

\noindent\makebox[\linewidth]{\rule{\columnwidth}{0.8pt}}
\textbf{Algorithm 1}: Redundant controlled operations removal

\vspace{-2mm}
\noindent\makebox[\linewidth]{\rule{\columnwidth}{0.4pt}}
\vspace{-6mm}
\begin{algorithmic}
\FORALL{$C[U]$ or $C^2[X]$ gate $g$ in the circuit}
    \STATE{execute circuit up to, but not including, $g$}
    \IF{$g$ is a $C[U]$ gate}
        \STATE{measure the control qubit $q$ in the $Z$ basis multiple times}
        \IF{$\{1\}$ is observed in the measurement results}
            \IF{$\{0\}$ is not observed in the measurement results}
                \STATE{turn $g$ into a $U$ gate acting on the target qubit}
            \ENDIF
        \ELSE
            \STATE{eliminate $g$}
        \ENDIF
    \ELSE
        \STATE{measure the control qubits $q_1 q_2$ in the $Z$ basis multiple times}
        \IF{$\{11\}$ is observed in the measurement results}
            \IF{neither $\{00\}$, $\{01\}$ nor $\{10\}$ is observed in the measurement results}
                \STATE{turn $g$ into an $X$ gate acting on the target qubit}
            \ELSIF{$\{01\}$ is not observed in the measurement results}
                \STATE{eliminate the control on $q_1$}
            \ELSIF{$\{10\}$ is not observed in the measurement results}
                \STATE{eliminate the control on $q_2$}
            \ENDIF
        \ELSE
            \STATE{eliminate $g$}
        \ENDIF
    \ENDIF
\ENDFOR
\end{algorithmic}
\vspace{-2mm}
\noindent\makebox[\linewidth]{\rule{\columnwidth}{0.4pt}}

%% file: application.tex
The \myopt optimization protocol described in Sec.~\ref{sec:algo} has been deployed to the quantum parton shower (QPS) algorithm~\cite{Nachman_2021}. We show experimental results from a simulator and quantum hardware, and discuss the optimization performance in terms of the number of CNOT gates and the identity of the final states.
The main purpose of this section is to demonstrate that \myopt optimization works with the determination of bitstrings by a noisy intermediate scale quantum computer in polynomial resources, not to compare with other initial-state dependent optimization protocols.

\subsection{Quantum parton shower algorithm}
\label{subsec:qalgo}
\input{app_algo}

\subsection{Experimental setup}
\label{subsec:setup}
\input{app_setup}

\subsection{Results}
\label{subsec:result}
\input{app_result}

%% file: app_algo.tex
The QPS algorithm in Ref.~\cite{Nachman_2021} can start with a fermion that is either type $f_1$ or $f_2$. These fermions can radiate a scalar particle $\phi$ or not at a given showering step. The relevant parameters are the three couplings $g_1$, $g_2$, and $g_{12}$ between $f_1$ and $\phi$, $f_2$ and $\phi$, and $f_1\bar{f}_2$ ($\bar{f}_1f_2$) and $\phi$, respectively, where the antifermion is denoted by $\bar{f}$. 
The shower evolution process is simulated by repeating the step by $N_\text{evol}$ times. When $N_\text{evol}$ is small, only a small number of particles are simulated, hence the number of non-zero computational basis states is also small.  In addition, since the circuit is designed to work with genetic initial states, a different set of computational basis states is occupied for each initial state, resulting in redundant controlled operations in the circuit.

\begin{figure}
\centering
\includegraphics[width=0.48\textwidth]{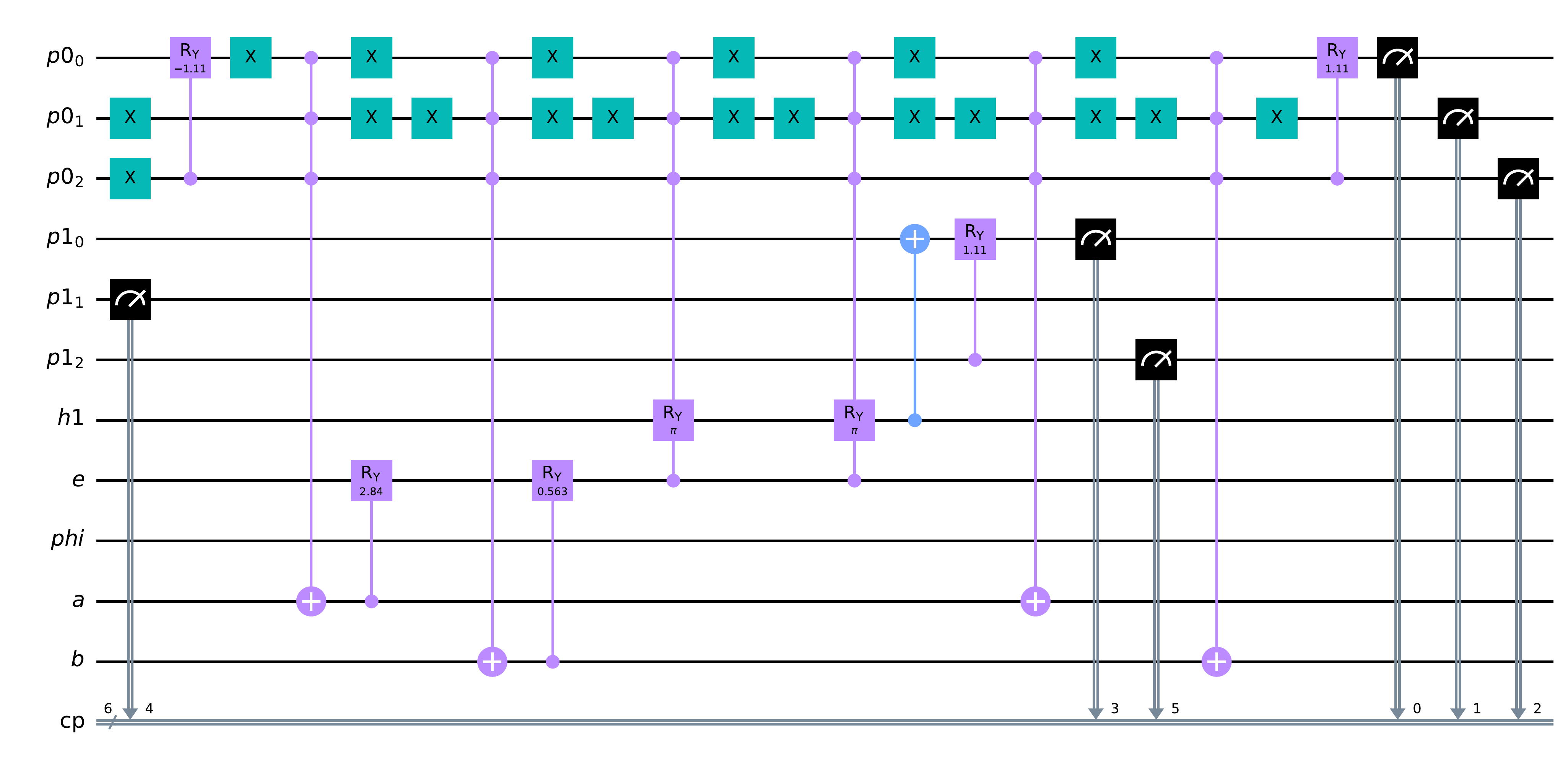}
\caption{Quantum circuit for the $N_\text{evol}=1$ step parton shower algorithm when the initial state is  $\ket{f_1}$. Coupling constants are $g_1=2$ and $g_2=g_{12}=1$.}
\label{fig:showercircuit}
\end{figure}

The quantum circuit for $N_\text{evol}=1$ step and the initial state $\ket{f_1}$ provide a good benchmark for the \myopt protocol. Coupling constants are set to $g_1=2$ and $g_2=g_{12}=1$.
Figure~\ref{fig:showercircuit} shows the benchmark quantum circuit for the QPS algorithm with $N_\text{evol}=1$.

%% file: app_setup.tex
\begin{figure}
\centering
\includegraphics[width=0.5\textwidth]{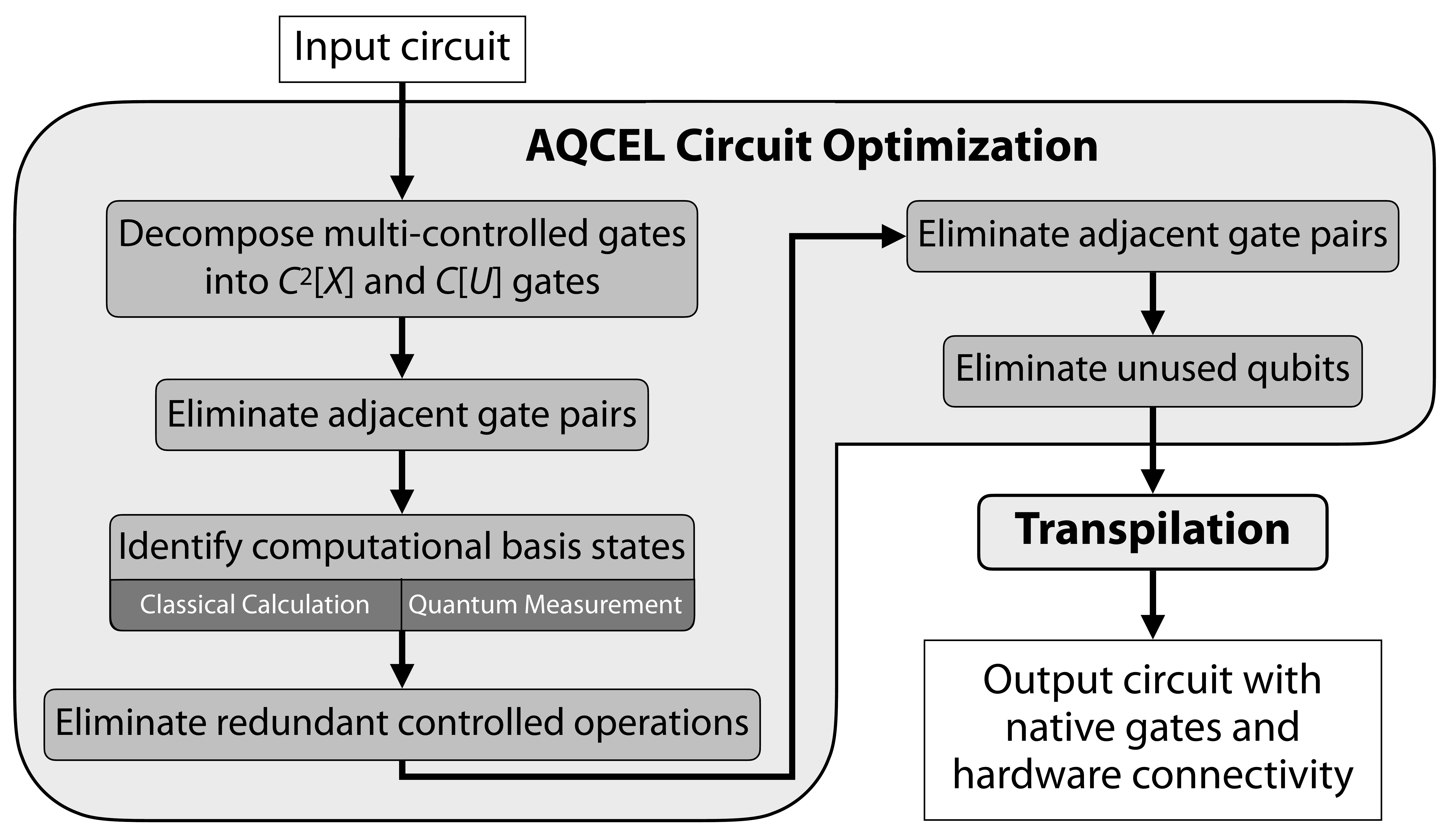}
\caption{Flowchart of the proposed optimization protocol. We eliminate unnecessary gates, qubit controls and unused qubits. Finally, the resulting circuit can be encoded into particular gates for specific hardware.}
\label{fig:flowchart} 
\end{figure}

The \myopt protocol focuses on circuit optimization at the algorithmic level, instead of at the level of a specific implementation using native gates for a particular quantum device. In addition to the initial-state dependent reduction of unnecessary controlled operations (Sec.~\ref{subsec:comp-base} and \ref{subsec:ctrl-gate}), the \myopt performs sequential decompositions of multi-controlled gates as well as the removal of adjacent gate pairs and unused qubits. 
A high-level flowchart of the \myopt protocol is shown in Fig.~\ref{fig:flowchart}.

The \myopt is implemented using IBM Qiskit version~0.32.1~\cite{Qiskit} with Terra~0.18.3, Aer~0.9.1 and Ignis~0.6.0 APIs in Python~3.8.1~\cite{10.5555/1593511}. The codes and experimental results are available in GitHub~\cite{code}. We attempt to optimize circuits running on a classical computer with a single 2.4~GHz Intel core i5 processor.
The \myopt optimization performance is evaluated using the 27-qubit IBM device called \textit{ibm\_kawasaki} equipped with the IBM Quantum Falcon Processor and the statevector simulator in Qiskit Aer. When executing a circuit on the \textit{ibm\_kawasaki}, the gates in the circuit are transformed into machine-native single- ($X, S_x, R_z$) and two-qubit (CNOT) gates, and the qubits are mapped to the hardware, accounting for the actual qubit connectivity of the \textit{ibm\_kawasaki}. For the results obtained solely from the statevector simulator, all the qubits are assumed to be connected to each other (referred to as the ideal topology) and the simulator does not consider any quantum noise.

In addition to \myopt, two circuit optimization tools are used: \tket in pytket~0.17.0 and pytket-qiskit~0.20.0, and IBM Qiskit transpiler. They are used as references for the comparison of the optimization performance as well as in the combination with \myopt to further reduce the gate counts.
For \tket, the {\it get\_compiled\_circuit} routine with optimization level 2 is used~\footnote{This routine is documented in the pytket manual at \texttt{https://cqcl.github.io/tket/pytket/api/backends.html}. The routine for decomposition all gates into basis gates is always applied in front of \tket.}. For Qiskit, the tranpilation with optimization level 3 pass manager is applied~\footnote{The value of {\it seed\_transpiler} is fixed to 1 in order to suppress the randomness of Qiskit transpiler.}.
The decomposition of multi-controlled gates using relative phase Toffoli gates is applied first to all the cases for comparison on an equal footing.

%% file: app_result.tex
Here we discuss results for the \myopt optimization to the $N_\text{evol}=1$ QPS circuit. The figure of merit is the numbers of single-qubit gates ($S_X$, $X$)~\footnote{$R_z$ gates are not included because they can be implemented with no cost as virtual Z gates~\cite{virtualZ}.} and CNOT gates obtained by decomposing quantum gates in the circuit and the calculation time of the circuit. The calculation time is defined as the duration of the pulse schedule of the transpiled circuit from the input to the measurements, as implemented in Qiskit.

\begin{figure}
\centering
\includegraphics[width=0.48\textwidth]{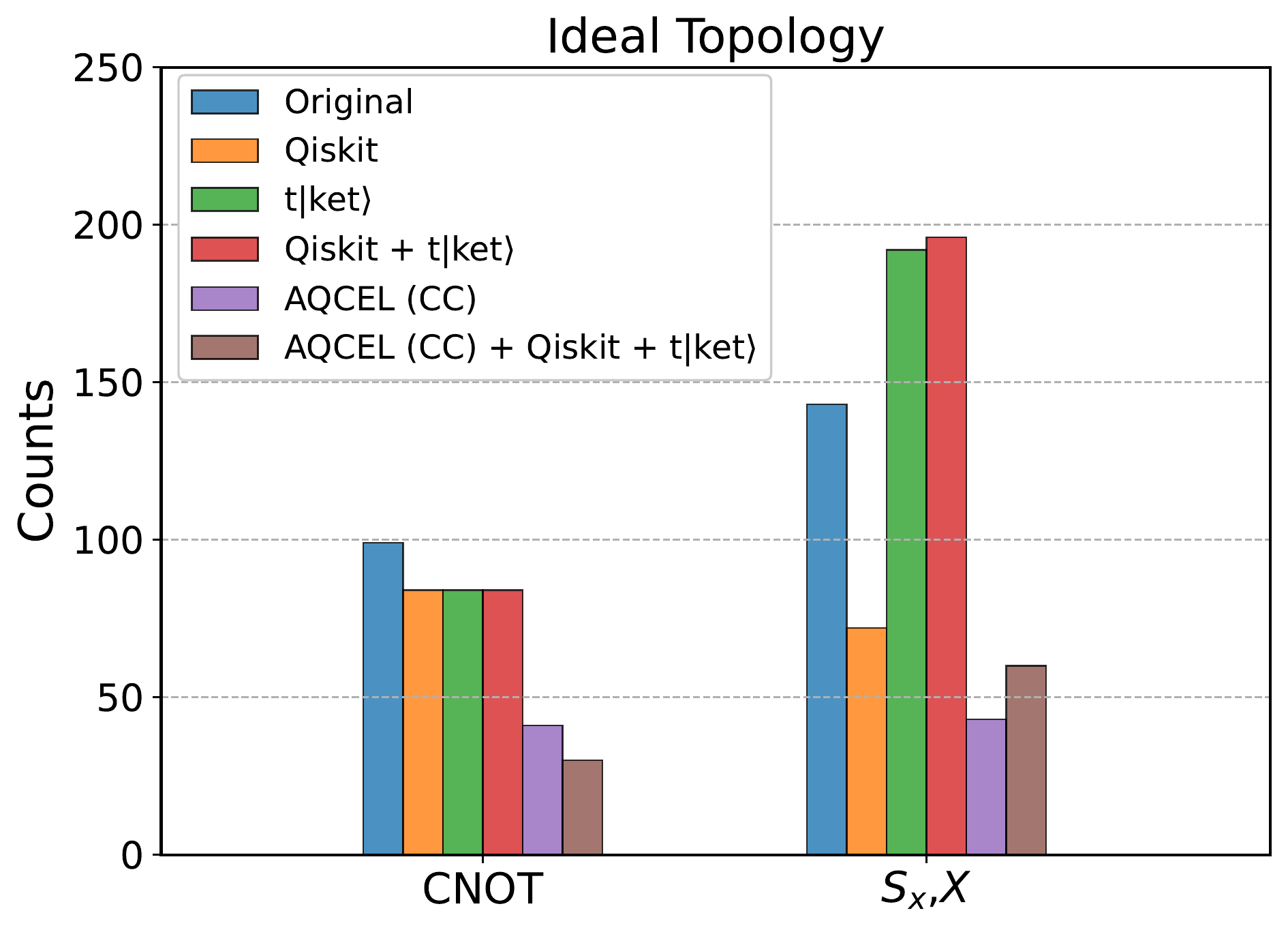}
\caption{Numbers of CNOT gates and single-qubit gates of the QPS circuit transpiled considering the ideal topology. For the \myopt optimization, the bitstring determination at controlled gates is performed using classical calculation.}
\label{fig:1step_ideal_gates} 
\end{figure}

First, we examine the numbers of single-qubit gates and CNOT gates assuming an ideal topology before and after the \myopt optimization alone. The determination of bitstrings at controlled gates is performed using classical calculation.
Figure~\ref{fig:1step_ideal_gates} shows the gate counts from the original circuit and the circuits optimized using either \myopt, \tket or Qiskit and their different combinations. It is seen that the \myopt alone reduces gate counts drastically, and even more for CNOT gate when the \myopt is combined with the \tket and Qiskit.

Starting with 155 CNOT and 431 single-qubit gates, the \myopt alone removes 114 CNOT and 388 single-qubit gates, in which 58 CNOT and 88 single-qubit gates are accounted for by the reduction of redundant qubit controls~\footnote{Removing redundant qubit controls reduces the cost of $C^2[X]$ and $C[U]$ decomposition into basis gates, resulting in the overall reduction of CNOT and single-qubit gate counts.}, and the rest by the removal of adjacent gate pairs.
The number of qubits is reduced from 14 to 13. The register $n_\phi$, composed of only one qubit, is removed because it is used only for the case where the initial state is $\ket{\phi}$.
The entire \myopt optimization takes about 0.58 seconds. The wall time is by far dominated by the elimination of redundant controlled operations (that accounts for 65\% of the total time), followed by a sub-dominant contribution of 35\% from the adjacent gate-pair elimination. 

\begin{figure}
\centering
\includegraphics[width=0.48\textwidth]{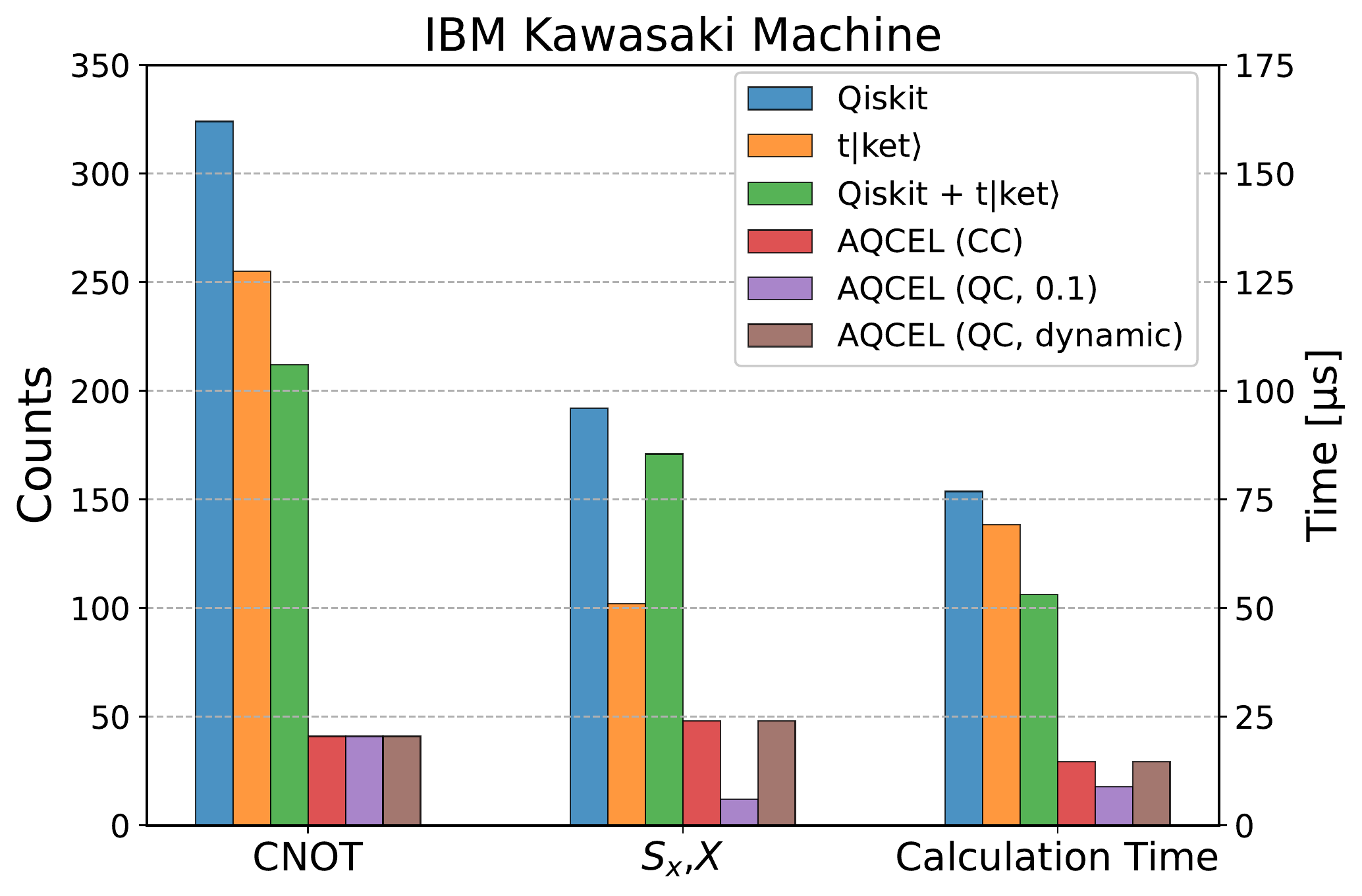}
\caption{Numbers of gates and the calculation times of the QPS circuit transpiled considering the topology of \textit{ibm\_kawasaki}. For the \myopt optimization, the bitstring determination at controlled gates is performed using either classical calculation (denoted by CC) or measuring the computational basis states on quantum hardware (denoted by QC) with the static thresholds of 0.1 or the dynamic threshold of \slow.}
\label{fig:1step_kawasaki_gates} 
\end{figure}

Now we evaluate the performance of the optimizers with the transpilation considering the hardware topology of the \textit{ibm\_kawasaki}. In \myopt, the bitstring determination is performed using classical calculation (denoted by CC) or quantum hardware (denoted by QC) with several thresholds. Figure~\ref{fig:1step_kawasaki_gates} shows the results for the $N_\text{evol}=1$ QPS circuit. The \myopt reduces the CNOT gate counts more significantly than the case for ideal topology (Fig.~\ref{fig:1step_ideal_gates}). This reduction comes from less SWAP gates (each of which requires 3 CNOTs) for the \myopt circuit because, once redundant qubit controls are removed, the amount of SWAP operations between multiple qubits is also suppressed. This results in much shorter calculation time for the \myopt circuit. For the rest in the paper, the most efficient transpilation that combines the \myopt, \tket and Qiskit is used for identification of the bitstrings at controlled gates and the fidelity measurement. The qubit counts are reduced from 14 to 13 under the dynamic threshold of \slow. Under the static thresholds, the qubit counts reduce to 13 for $0.005 \leq \sfrac \leq 0.25$, but more significantly to 8 when $\sfrac = 0.3$.

In the initial-state dependent circuit optimization, what is preserved is the equivalence of the final state, not the circuit itself. To evaluate the accuracy of the \myopt optimization, we consider a classical fidelity between final states before and after the optimization, defined in terms of the probability distributions of the bitstrings observed in the measurement at the end of the circuits. This quantity, denoted as $F$ and referred to as just ``fidelity'' hereafter, is given by
\begin{equation}
F = \sum_k \sqrt{\porig_k \popt_k},
\end{equation}
where the index $k$ runs over the bitstrings. The quantities $\porig_k$ and $\popt_k$ are the probabilities of observing $k$ in the original and optimized circuits, respectively.

In fact, we compute two fidelity values for each optimization method. For the ideal final state where any quantum error is not considered, the first fidelity, denoted $\Fsim$, aims to quantify the amount of modifications to the final state introduced by the optimization procedure at the algorithmic level. To calculate the $\Fsim$, both {\porig} and {\popt} are computed using the statevector simulation. The value of $\Fsim = 1$ indicates that the final states are identical before and after the optimization (up to a possible phase difference on each of the qubits), while a deviation from unity gives a measure of how much the optimization has modified the final state from the ideal one.

The second fidelity value, $\Fmeas$, is computed using measurements with an actual quantum computer for $\popt$. The $\popt$ is estimated from the rate at which a bitstring occurs in a large number of repeated measurements. The {\porig} is computed using simulation, as for the $\Fsim$. 
Even if $\Fsim$ is 1, the presence of noise will make $\Fmeas < 1$, with the difference from unity getting larger when more gates (particularly CNOT gates) are present in the circuit. Removing CNOT gates to optimize the circuit will lower the overall effect of noise and raise the $\Fmeas$ value. However, when low-amplitude computational basis states are rejected by the threshold, more qubit controls are removed, which makes the final state different from the ideal one and decreases the {\Fmeas} value. Thus, the {\Fmeas} is a measure that reflects the trade-off of making the circuit shorter and changing the final state through the optimization. The measurements are performed 10000 times for each optimized circuit to obtain the {\Fmeas} value, and the experiment is repeated 30 times with the same optimized circuit to finally obtain the average and the standard deviation of the {\Fmeas} values. Any error mitigation is not used in the measurements.

\begin{figure}
\centering
\includegraphics[width=0.48\textwidth]{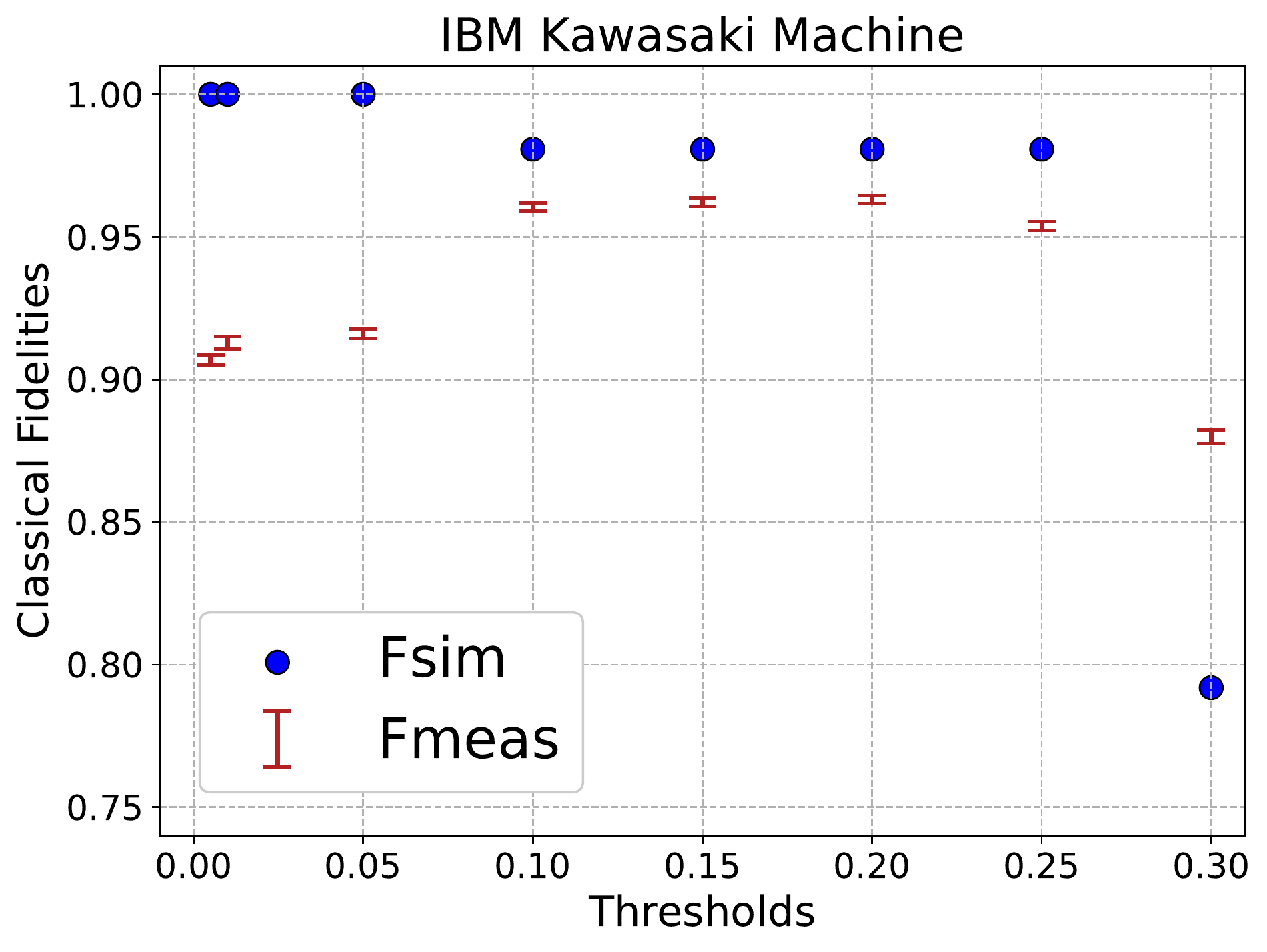}
\caption{{\Fsim} and {\Fmeas} values as a function of the static threshold \sfrac used to optimize the QPS circuits by \myopt. The bitstring measurement on controlled qubits is performed on the \textit{ibm\_kawasaki}.}
\label{fig:fidelity_constant} 
\end{figure}

When the elimination of redundant qubit controls is performed based on measurements using a quantum computer with the static thresholds \sfrac, the threshold dependence of {\Fsim} and {\Fmeas} values is shown in Fig.~\ref{fig:fidelity_constant}. 
With increasing \sfrac value the $\Fmeas$ first increases, indicating the suppression of noise effects due to CNOT gate removal, but then worsens significantly at $\sfrac=0.30$. This is understood from the behavior of the {\Fsim} value: the {\Fsim} stays close to unity up to $\sfrac=0.25$ then decreases significantly, signaling that the optimization is too aggressive to maintain the final state from the original one if $\sfrac=0.30$ is used. For the circuit considered here, the performance of the optimization appears to be best with $0.10 \leq \sfrac \leq 0.25$. 

\begin{figure}
\centering
\includegraphics[width=0.48\textwidth]{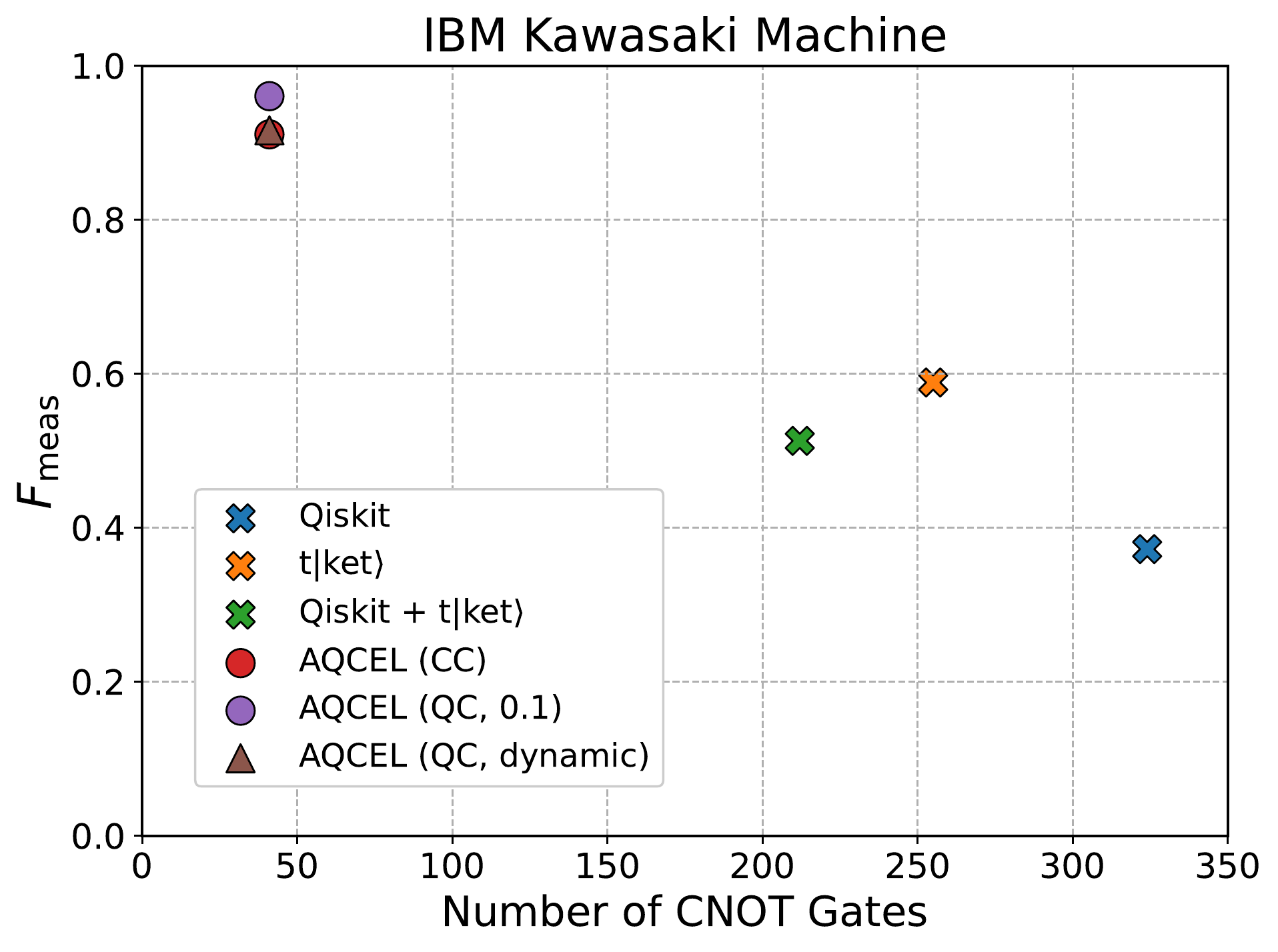}
\caption{Fidelity {\Fmeas} versus the number of CNOT gates for the QPS circuit transpiled considering the topology of
\textit{ibm\_kawasaki}. For the \myopt optimization, the bitstring determination at controlled gates is performed
using classical calculation (denoted by CC) or measuring the computational basis states on quantum hardware (denoted by QC) with the static thresholds of 0.1 or the dynamic threshold of \slow.}
\label{fig:fidelity_1step_ps} 
\end{figure}

Shown in Fig.~\ref{fig:fidelity_1step_ps} is the {\Fmeas} versus CNOT gate counts with the determination of bitstrings using classical calculation and hardware measurement.
For the \slow, the optimized circuit turns out to be exactly same as the one obtained using classical calculation~\footnote{Although these circuits are identical, the {\Fmeas} values are slightly different due to statistical uncertainty and the quantum noise of the actual device.}. 
For the \myopt optimized circuits, the {\Fmeas} values are $0.961\pm0.001$ for $\sfrac=0.1$ and $0.911\pm0.002$ for the optimization using classical calculation. This demonstrates a clear improvement from the hardware measurement by removing qubit controls that trigger on low-amplitude computational basis states. The {\Fsim} value is 0.981 for $\sfrac=0.1$, meaning that the final state is modified slightly from the ideal one.

To further evaluate the accuracy of the \myopt optimized circuit, a quantum state tomography (QST) is performed over the particle registers of six qubits in the $N_{\rm evol}=1$ QPS circuit (see Fig.~\ref{fig:showercircuit}). From the measurements of $3^6$ circuits, each performed 4000 times, with Pauli $\{X,Y,Z\}$ observables using the quantum hardware, the density matrix is reconstructed for the optimized circuit and compared with one obtained from the original circuit to compute a fidelity (denoted by {\Fqst}). In a statevector simulation, the {\Fqst} value is unity for both Qiskit+\tket and \myopt circuit with the \slow, meaning that the optimization does not modify the final state including relative phases. The {\Fqst} values are measured to be 0.13 and 0.41 for Qiskit+\tket and \myopt circuits, respectively, when performing QST on the quantum hardware. The {\Fqst} values are much smaller than unity due to hardware noise, but the \myopt shows a clear improvement over the Qiskit+\tket.





%% file: discussion.tex
\subsection{Applicability of proposed optimization}
\label{diss:applicability}
The core component of the proposed circuit optimization is the identification of computational basis states with zero- or low-amplitudes using a quantum computer and the subsequent elimination of redundant controlled operations. Therefore, the \myopt is expected to work more efficiently for quantum algorithms in which the quantum state has a small number of high-amplitude computational basis states. In other words, the \myopt would not be effective if all the computational basis states have non-negligible amplitudes, especially when they are small in size because of thresholds applied on quantum hardware. A typical example is Quantum Phase Estimation~\cite{nielsen00} or Grover's Algorithm, where an equal superposition state is created first by applying $H^{\otimes n}$ gates to the initial state $\ket{0}^{\otimes n}$ of the $n$-qubit system. However, even in this case, the \myopt can efficiently produce a quantum circuit that approximates the original final state by ignoring low-amplitude computational basis states.

Another important aspect for the \myopt optimization protocol is the resource needed to use a quantum computer for the optimization. In the NISQ era, it is worth spending the quantum computer resource in the optimization if the resulting circuit can produce higher-fidelity results than the original circuit does. In the fault-tolerant quantum computing era, using a quantum computer for the optimization may not be crucial due to its capability of correcting quantum errors during the operation. However, the initial-state dependent \myopt optimization will be still useful for simplifying the quantum circuit, even in the fault-tolerant regime.

\subsection{Further simplifications with initial-state dependent optimization}
\label{diss:gate_reduce}
The new method proposed here for obtaining bitstrings using a quantum computer will open new possibilities in initial-state dependent quantum circuit optimization within polynomial computational resources. In this paper, we discuss the simplest example of this type of optimizations, that is, just use the information of control qubits and optimize multi-controlled qubit gates individually. There are several possibilities to extend the idea for further simplifications, e.g., using ancilla qubits, quantum gates with qutrit (3-level) states, or adding the information of target qubit in controlled gate like RPO~\cite{liu2020relaxed}. One can optimize not only individual gates but also multiple gates as a gate set in future.

As another interesting possibility, if a circuit turns out to contain only a small number of basis states, one could represent the circuit state using fewer qubits than the original ones. Given that this approach might require a completely new computational basis, this is left for future work.

\subsection{Mitigations of quantum errors}
\label{diss:error_mitigation}
The threshold choice in \myopt has significant impacts on {\Fsim} and \Fmeas, as seen in Figs.~\ref{fig:fidelity_constant} and \ref{fig:fidelity_1step_ps}. 
The measurement error can be improved by adapting the unfolding technique developed in Ref.~\cite{Bauer:2019uf} and related approaches that use fewer resources~\cite{geller_efficient_2020,song_10-qubit_2017,gong_genuine_2019,wei_verifying_2020,hamilton2020scalable} or further mitigate the errors~\cite{2010.07496}. 
A substantial contribution to the gate errors originates from CNOT gates. There are a variety of approaches to mitigate these errors, including the zero noise extrapolation with identity insertions, first proposed in Ref.~\cite{Dumitrescu:2018} and generalized in Ref.~\cite{PhysRevA.102.012426}.
The method based on the CNOT error mitigation may improve the accuracy of optimizations and the fidelity of our approach.

%% file: conclusion.tex
We have proposed a new optimization protocol, called \myopt, for analyzing quantum circuits to remove redundant controlled operations. The \myopt can remove unnecessary qubit controls from multi-controlled gates even when all the relevant qubits are entangled. The heart of the redundant controlled operations removal resides in the identification of zero- or low-amplitude computational basis states. In particular, this procedure can be performed through measurements using a quantum computer in polynomial time,
instead of classical calculation that scales exponentially with the number of qubits. Although removing qubit controls that trigger on low-amplitude basis states will result in a circuit that produces the final state distinct from the original one, this may be a desirable feature under the existence of hardware noise.

We have adopted the \myopt optimization scheme to the quantum parton shower simulation using the \textit{ibm\_kawasaki}.
In the experiment, the proposed scheme has shown a significant reduction of gate counts and improved the {\Fmeas} value while retaining the accuracy of the probability distributions of the final state.
For the \myopt optimized circuits, the {\Fmeas} values are $0.961\pm0.001$ for the static threshold $\sfrac=0.1$ and $0.911\pm0.002$ for the optimization using classical calculation. For the dynamic threshold \slow, the optimized circuit is exactly same as the one obtained using classical calculation.
The initial-state dependent optimization discussed here opens new possibilities to extend quantum circuit optimization further in future.

%% file: acknowledgements.tex
We acknowledge the use of IBM Quantum Services for this work. The views expressed are those of the authors, and do not reflect the official policy or position of IBM or the IBM Quantum team.

This study is partly carried out under the project ``Optimization of HEP Quantum Algorithms'' supported by the U.S.-Japan Science and Technology Cooperation Program in High Energy Physics.

CWB and BN are supported by the U.S. Department of Energy, Office of Science under contract DE-AC02-05CH11231.  In particular, support comes from Quantum Information Science Enabled Discovery (QuantISED) for High Energy Physics (KA2401032).

This research used resources of the Oak Ridge Leadership Computing Facility, which is a DOE Office of Science User Facility supported under Contract DE-AC05-00OR22725.

We would like to thank Ross Duncan and Bert de Jong for useful discussions about the ZX-calculus.

%% file: appendix.tex
\section{Decomposition of multi-controlled gates}
\label{app:decomposition}
When the same qubit controls are removed from $C^{m}[U]$ and $C^{s}[U] (m>s)$, Eq.~\eqref{eq:aqcel_condition} indicates that the condition for the removal of qubit controls is stricter for the latter because the number of $k$ is larger. This means that if we decompose a $C^{m}[U]$ into controlled gates with smaller number of control qubits, e.g., Toffoli and two-qubit gates, the opportunity to remove redundant qubit controls is partly lost. This suggests that removing redundant qubit controls should be applied before decomposing multi-controlled gates into basis gates like CNOT.
As mentioned in Sec.~\ref{subsec:qubit_control} and Appendix~\ref{app:comp_resource}, the decomposition of multi-controlled gates is essential in terms of computational resources. This means that there is a trade-off between the opportunities of removing qubit controls and computational complexity.

\section{Computational resources for the proposed optimization scheme}
\label{app:comp_resource}

The computational costs to perform the proposed optimization scheme are evaluated here.
We consider a quantum circuit that contains $n$ qubits and $N$ multi-qubit controlled gates, each acting on $m$ control qubits and one target qubit.

The first step in the optimization scheme is the identification of computational basis states.
If we use the classical calculation to simply track all the computational basis states whose amplitudes may be nonzero at each point of the circuit, 
it requires the computation of ${\cal O}(N2^n)$ states which grows exponentially with $n$. This method requires less computational resource than a statevector simulation, but it neglects certain rare cases where exact combinations of amplitudes lead to the elimination of redundant controlled operations. This is because a statevector simulation is applied as a classical calculation for the identification of bitstrings in \myopt.
If we measure the control qubits at each controlled gate $M$ times using a quantum computer, the total number of gate operations and measurements is given by
\begin{multline}\label{eq:the_condition}
  M\{m+(1+m)+(2+m)+\cdots+(N-1+m)\} \\
  = \frac{1}{2}MN(N-1)+mMN
\end{multline}
Therefore, the computational cost grows polynomially with $n$ in ${\cal O}(MN^2+mMN)$.

We next consider the decision of all redundant qubit controls from a controlled gate with $m$ control qubits.
Using a quantum computer that measures the $m$ control qubits $M$ times, the measured number of bitstrings is $M$. For the classical calculation, the number of basis states is $2^m$.
Imagine that we choose an arbitrary combination among $2^m$ possible combinations of new qubit controls on the same controlled gate. We need to check if all specified bitstrings in Eq.~\eqref{eq:aqcel_condition} for the chosen combination are not measured. The cost is ${\cal O}(Mm2^m)$ for one chosen combination because the size of the bitstring is $m$ and the numbers of measured and specified bitstrings are $M$ and $2^m$, respectively. Therefore, the overall computational cost for the determination of redundant qubit controls is ${\cal O}(Mm4^mN)$ for $N$ multi-qubit controlled gates. The classical calculation requires ${\cal O}(m8^mN)$ as well.

In the \myopt protocol, all controlled gates in the circuit are decomposed into Toffoli gates and two-qubit controlled-$U$ gates. With this decomposition, the total number of gate operations and measurement increases due to ${\cal O}(m)$ times more controlled gates. However, the computational cost for the redundant qubit control identification becomes polynomial in ${\cal O}(MmN)$ because all controlled gates have the constant number of control qubits ($m=1,2$). The computational cost for the identification of computational basis states still behaves polynomially in ${\cal O}(m^2MN^2)$ when a quantum computer is used. Given that a controlled gate has at most $n-1$ control qubits, the total computational cost for the entire optimization sequence is ${\cal O}(n^2MN^2)$, when the computational basis state measurement is performed using a quantum computer.